\documentclass[letterpaper,english]{IEEEtran}
\usepackage[T1]{fontenc}
\usepackage[latin9]{inputenc}
\usepackage{amsmath}
\usepackage{graphicx}
\usepackage{amssymb}

\makeatletter

\newcommand{\lyxline}[1][1pt]{%
  \par\noindent%
  \rule[.5ex]{\linewidth}{#1}\par}

\newtheorem{thm}{Theorem}
\newtheorem{remrk}{Remark}
\newtheorem{example}{Example}

\usepackage{cite}
\usepackage{times}
\usepackage{hyperref}
\usepackage{babel}
\makeatother

\begin{document}

\title{A Simple Converse Proof and a Unified Capacity Formula for Channels
with Input Constraints}

\author{{\normalsize Youjian (Eugene) Liu}%
\thanks{This work was supported by NSF Grants CCF-0728955, ECCS-0725915, and
Thomson Inc. A single column version of this paper was submitted to
IEEE Transactions on Information Theory on June 17, 2008.%
}{\normalsize \\Department of Electrical and Computer Engineering\\University
of Colorado at Boulder\\eugeneliu@ieee.org}}

\maketitle
\begin{abstract}
Given the single-letter capacity formula and the converse proof of
a channel without input constraints, we provide a simple approach
to extend the results for the same channel but with input constraints.
The resulting capacity formula is the minimum of a Lagrange dual function.
It gives an unified formula in the sense that it works regardless
whether the problem is convex. If the problem is non-convex, we show
that the capacity can be larger than the formula obtained by the naive
approach of imposing constraints on the maximization in the capacity
formula of the case without the constraints. 

The extension on the converse proof is simply by adding a term involving
the Lagrange multiplier and the constraints. The rest of the proof
does not need to be changed. We name the proof method the Lagrangian
Converse Proof. In contrast, traditional approaches need to construct
a better input distribution for convex problems or need to introduce
a time sharing variable for non-convex problems. We illustrate the
Lagrangian Converse Proof for three channels, the classic discrete
time memoryless channel, the channel with non-causal channel-state
information at the transmitter, the channel with limited channel-state
feedback. The extension to the rate distortion theory is also provided.
\end{abstract}

\section{Introduction\label{sec:Introduction}}

Naively imposing input constraints on the maximization in the single-letter
capacity formula of a channel without input constraints often produces
the capacity formula of the same channel with the constraints. For
example, the classic discrete time memoryless channel without input
constraints has capacity \begin{eqnarray*}
C' & = & \max_{p_{X}}I(X;Y),\end{eqnarray*}
 and with a power constraint, the capacity is \begin{eqnarray*}
C & = & \max_{p_{X}:\textrm{E}X^{2}\le\rho_{0}}I(X;Y).\end{eqnarray*}
 Such cases are so prevalent that one may suspect it is always the
case. We started with this belief while working on channels with limited
channel-state feedback. If one denotes the single letter capacity
for the case without constraint as \begin{eqnarray}
C' & = & \max\textrm{ Mutual Information},\label{eq:capacity-text}\end{eqnarray}
 contrary to the conventional belief, we found in \cite{Liu_Asi07_Capacity_thms_feedback,Liu_IT07s_Capacity_MAC}
that the capacity for the case with the constraint can be larger than
\begin{eqnarray*}
R & = & \max_{\mbox{constraint}}\textrm{ Mutual Information}\end{eqnarray*}
and the capacity can be expressed as \begin{eqnarray}
C & = & \min_{\lambda\ge0}\textrm{ Lagrange Dual Function}(\lambda)\label{eq:cap-w-const-text}\\
 & = & R+\textrm{Duality Gap}\nonumber \\
 & \ge & R,\label{eq:cap-lg-R}\end{eqnarray}
 where the Lagrange dual function \cite{Boyd_04Book_Convex_optimization}
to the primary problem $R$ counts for the constraint. 

Capacity formula (\ref{eq:cap-w-const-text}) reduces to the maximum
of the mutual information when the duality gap is zero and therefore,
Equation (\ref{eq:cap-w-const-text}) is an unified expression for
cases with non-zero or zero duality gaps. 

During the discovery of the capacity result for the channel with limited
feedback and with constraints, we found a new proof of the converse
part of the capacity theorem. It is obtained via modifying the converse
proof for the case without the constraints by adding to the second
to the last expression a term involving the Lagrange multiplier and
the constraints. The rest of the proof is unchanged. We call such
a proof the \emph{Lagrangian Converse Proof}. With little modification,
the method can also be used to prove the converse part of the rate
distortion theorem. The unexpected simplicity and the potential to
obtain new results with ease motivates us to report it here.

A meaningful theory should be able to explain the past and predict
the future. In this paper, we show that the Lagrangian Converse Proof
can simplify the existing proof of the capacity of the classic discrete
memoryless channels and the proof of the capacity of the channels
with non-causal channel-state information at the transmitters (CSIT)
\cite{Gelfand:80,Cover_02IT_Duality_two_sided_CSI,Sullivan_IT03_Information_hiding}.
In addition, we illustrate how to use it to obtain new capacity results
of the channels with limited channel-state feedback \cite{Liu_Asi07_Capacity_thms_feedback,Liu_IT07s_Capacity_MAC}. 

To understand why the capacity can be greater than the maximum of
the mutual information as shown in (\ref{eq:cap-lg-R}), we provides
a convex hull explanation of the capacity region of the single user
channel. Yes, even for single user channels, investigating the capacity
region is meaningful when the capacity needs to be achieved using
time sharing. The minimum of the Lagrange dual function conveniently
characterize the capacity region's boundary points without explicitly
employing the time sharing. The intuition is that when the duality
gap is greater than zero, multiple solutions to (\ref{eq:cap-w-const-text})
exist. Some solution is below the constraint and some is above the
constraint. A time sharing of the solutions will achieve the capacity
and at the same time, satisfy the constraint exactly. Therefore, the
capacity can alternatively be expressed as the maximum of the time
sharing of the mutual information. 

In summary, the contributions of the paper are as follows.

\begin{itemize}
\item A simple converse proof is provided for the capacities of channels
with constraints and for rate distortion theorems;
\item Expressed using the Lagrange dual function, an unified capacity formula
is presented and shown to have an intimate relation to the convex
hull of the capacity region and the time sharing. Free of time sharing
variables, the expression also makes the calculation of the capacities
easier. The capacity formula also has a pleasant symmetric relation
to rate distortion function.
\end{itemize}

In Section \ref{sec:Capacity-Theorems}, the simplicity of the Lagrangian
converse proof is illustrated for three channels, the discrete memoryless
channel, the channel with non-causal channel-state information, and
the channel with limited channel-state feedback. For the latter, the
relation among the capacity formula, the capacity region, and the
time sharing is explained. In Section \ref{sec:rate-distortion},
the converse proof is extended to the rate distortion theory. The
dual relation of channel capacity and rate distortion is briefly discussed.
Section \ref{sec:Conclusions} summarizes the usage of the proposed
converse proof.

\section{The Lagrangian Converse Proof for Channel Capacities\label{sec:Capacity-Theorems}}

There are two traditional methods of converse proof for channels with
input constraints. The first method takes advantage of the convexity
of the problem and produces a better input distribution from any input
distribution induced by the information message and the code. This
better input distribution must also satisfy the input constraints.
Section \ref{sub:DMC} compares this method with the Lagrangian Converse
Proof for the classic discrete memoryless channels. The second method
is to introduce a time sharing variable for non-convex problems. Section
\ref{sub:non-causal} and \ref{sub:feedback} compares it with the
new converse proof for channels with non-causal channel-state information
at the transmitter and for channels with limited feedback, of which
an example of nonzero duality gap is provided.

\subsection{\label{sub:DMC}The Capacity of the Discrete Memoryless Channels}

\begin{figure}
\begin{centering}
\textsf{\includegraphics[clip,angle=270,scale=0.3]{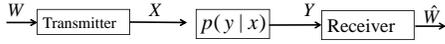}}
\par\end{centering}

\caption{\label{cap:cha-DMC} A discrete memoryless channel}

\end{figure}

The channel $(\mathcal{X},p_{Y|X},\mathcal{Y})$ in Figure \ref{cap:cha-DMC}
is a memoryless channel with finite alphabets $(\mathcal{X},\mathcal{Y})$
for input $X\in{\mathcal{X}}$ and output $Y\in{\mathcal{Y}}$. The
inputs over $N$ channel satisfy the constraint, \begin{equation}
\frac{1}{N}\sum_{n=1}^{N}\textrm{E}\left[\alpha(X_{n})\right]\le\rho_{0},\label{eq:constraint-no-feedback}\end{equation}
 where the expectation is over the information message and $\alpha(\cdot):\mathcal{X}\rightarrow\mathbb{R}$
is a real valued function. For example, it is a power constraint if
$\alpha(X)=X^{2}$. 

It is well known that the capacity of this channel without the constraint
is \begin{eqnarray}
C'_{1} & = & \max_{p_{X}}I(X;Y),\label{eq:capacity-DMC}\end{eqnarray}
 and with the constraint, the capacity is\begin{eqnarray}
R_{1} & = & \max_{\begin{array}{c}
p_{X}:\textrm{E}[\alpha(X)]\le\rho_{0}\end{array}}I(X;Y).\label{eq:primary-no-feedback}\end{eqnarray}
 The Lagrange dual function of (\ref{eq:primary-no-feedback}) is
\begin{eqnarray}
L_{1}(\lambda,\rho_{0}) & \triangleq & \max_{p_{X}}I(X;Y)-\lambda(\textrm{E}[\alpha(X)]-\rho_{0}),\label{eq:Lag-no-feedback}\end{eqnarray}
 which is an upper bound to $R_{1}$ for all $\lambda\ge0$ and all
$p_{X}$ that satisfy the constraint $\textrm{E}[\alpha(X)]\le\rho_{0}$
\cite{Boyd_04Book_Convex_optimization}. The duality gap $G_{1}$
is defined as the least upper bound minus $R_{1}$, i.e., \begin{eqnarray*}
G_{1} & = & \inf_{\lambda\ge0}L_{1}(\lambda,\rho_{0})-R_{1}.\end{eqnarray*}
 Because the mutual information is a convex $\cap$ function of the
input distribution $p_{X}$ and the input constraint is convex, $R_{1}$
is a convex $\cap$ function of $\rho_{0}$. Therefore, the duality
gap $G_{1}$ is zero \cite{Yu_06_TCOM_Dual_nonconvex_opt} and the
capacity can be expressed as \begin{eqnarray}
C_{1} & = & \min_{\lambda\ge0}L_{1}(\lambda,\rho_{0})\label{eq:C-Finite-no-feedback}\\
 & = & L_{1}(\lambda^{*},\rho_{0})=R_{1}.\label{eq:lmd-star-no-feedback}\end{eqnarray}

We compare the converse proof with and without the constraint. The
last step of the converse proof for the case without the constraint
is \begin{eqnarray*}
\sum_{n=1}^{N}I(X_{n};Y_{n}) & \le & NC_{1}^{'},\end{eqnarray*}
 where $C_{1}^{'}$ dominates $I(X_{n};Y_{n})$ for \emph{every} $n$.
With input constraint the additional steps of the \emph{traditional}
proof of the converse \cite[Chapter 7.3]{Gallager:68} are \begin{eqnarray}
\sum_{n=1}^{N}I(X_{n};Y_{n}) & \le & NI(X;Y)\label{eq:old-get-input}\\
 & \le & NR_{1},\label{eq:old-R0}\end{eqnarray}
 where, unlike the case without input constraint, $R_{1}$ may not
dominate every $I(X_{n};Y_{n})$ because the constraint (\ref{eq:constraint-no-feedback})
is averaged over $N$ channel uses and thus it is possible that $\textrm{E}[\alpha(X_{n})]>\rho_{0}$
for some $n$. One has to construct a new input distribution $P_{X}(x)=\frac{1}{N}\sum_{n=1}^{N}p_{X_{n}}(x)$
and use the property that the mutual information is a convex $\cap$
function of input distribution to obtain (\ref{eq:old-get-input}).
Luckily, the new input distribution satisfies the constraint $\textrm{E}[\alpha(X)]\le\rho_{0}$
because $\textrm{E}[\alpha(X)]$ is a convex function of $p_{X}$,
and thus one obtains (\ref{eq:old-R0}).

Using the Lagrangian Converse Proof, the key step is to add a term
of Lagrange multiplier: \begin{eqnarray}
 &  & \sum_{n=1}^{N}I(X_{n};Y_{n})\nonumber \\
 & \le & \sum_{n=1}^{N}\left(I(X_{n};Y_{n})-\lambda^{*}\left(\textrm{E}\left[\alpha(X_{n})\right]-\rho_{0}\right)\right)\label{eq:new-summand}\\
 & \le & NC_{1},\label{eq:new-C0}\end{eqnarray}
 where $\lambda^{*}\ge0$ is the solution in (\ref{eq:lmd-star-no-feedback});
(\ref{eq:new-summand}) follows from the fact that $X_{n}$'s satisfy
the constraint and thus $-\lambda^{*}\left(\left(\sum_{n=1}^{N}\textrm{E}\left[\alpha(X_{n})\right]\right)-N\rho_{0}\right)\ge0$;
(\ref{eq:new-C0}) follows from the fact that $C_{1}$ of (\ref{eq:lmd-star-no-feedback})
dominates the summand in (\ref{eq:new-summand}) for \emph{every}
$n$, as in the case without constraints, because the power penalty
$\lambda^{*}\left(\textrm{E}\left[\alpha(X_{n})\right]-\rho_{0}\right)$
punishes excessive power use. The simplification is that we do not
need to construct a better input distribution $p_{X}$. It will be
significant when there is no obvious way to find a better $p_{X}$.

\subsection{\label{sub:non-causal}The Capacity of Channels with Non-causal Channel-state
Information at the Transmitter}

\begin{figure}
\begin{centering}
\textsf{\includegraphics[clip,angle=270,scale=0.3]{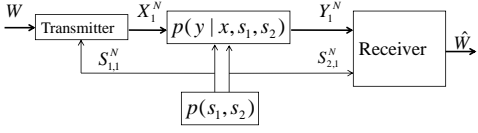}}
\par\end{centering}

\caption{\label{cap:cha-non-causal} A channel with non-causal channel-state
information at the transmitter}

\end{figure}

As shown in Figure \ref{cap:cha-non-causal}, the memoryless channel
with finite alphabets is characterized by $(\mathcal{X},\mathcal{Y},\mathcal{S}_{1},\mathcal{S}_{2},p_{S_{1},S_{2}},p_{Y|X,S_{1},S_{2}})$,
where $X\in{\mathcal{X}}$ is the channel input; $(S_{1},S_{2})\in{(\mathcal{S}_{1},\mathcal{S}_{2})}$
is the channel-state with distribution $p_{S_{1},S_{2}}$; $p_{Y|X,S_{1},S_{2}}$
is the channel transition probability; and $(Y\in{\mathcal{Y}},S_{2}\in{\mathcal{S}_{2}})$
is the channel output, i.e., the channel-state $S_{2}$ is non-causally
known at the receiver. The channel-state $S_{1}$ is non-causally
known at the transmitter. In the proof of the converse, the inputs
over $N$ channel uses satisfy the constraint, \begin{equation}
\frac{1}{N}\sum_{n=1}^{N}\textrm{E}\left[\alpha(X_{n})\right]\le\rho_{0},\label{eq:constraint-noncausal}\end{equation}
 where the expectation is over the information message and the state
$S_{1}$.

Without input constraints, the capacity is directly obtained in \cite{Cover_02IT_Duality_two_sided_CSI}
or can be obtained from \cite{Gelfand:80} by considering $(Y\in{\mathcal{Y}},S_{2}\in{\mathcal{S}_{2}})$
as the channel output. The capacity is \begin{eqnarray*}
C_{2}^{'} & = & \max_{\mathcal{U},X=\varphi(U,S_{1}),p_{U|S_{1}}}I(U;S_{2},Y)-I(U;S_{1}),\end{eqnarray*}
 where $X$ is a deterministic function of $U$ and $S_{1}$, $U\in\mathcal{U}$
is an auxiliary random variable. 

With the input constraint, the capacity is \begin{eqnarray}
C_{2} & = & \min_{\lambda\ge0}L_{2}(\lambda,\rho_{0})\label{eq:C-Finite-noncausal}\\
 & = & L_{2}(\lambda^{*},\rho_{0})\label{eq:lmd-star-noncausal}\\
 & = & R_{2},\label{eq:NC2}\end{eqnarray}
 where \begin{eqnarray}
R_{2} & = & \max_{\mathcal{U},X=\varphi(U,S_{1}),p_{U|S_{1}}:\textrm{E}[\alpha(X)]\le\rho_{0}}\nonumber \\
 &  & I(U;S_{2},Y)-I(U;S_{1});\label{eq:R2-noncausal}\end{eqnarray}
\begin{eqnarray}
L_{2}(\lambda,\rho_{0}) & \triangleq & \max_{\mathcal{U},X=\varphi(U,S_{1}),p_{U|S_{1}}}I(U;S_{2},Y)-I(U;S_{1})\nonumber \\
 &  & \qquad\qquad-\lambda(\textrm{E}[\alpha(X)]-\rho_{0})\},\label{eq:dual-function-noncausal}\end{eqnarray}
 is the Lagrange dual function to the primary problem (\ref{eq:R2-noncausal});
\begin{eqnarray*}
\textrm{E}[\alpha(X)] & = & \sum_{s_{1}}\sum_{u}p_{S_{1}}(s_{1})p_{U|S_{1}}(u|s_{1})\alpha\left(\varphi(u,s_{1})\right);\end{eqnarray*}
 (\ref{eq:NC2}) follows from the fact that $U$ can include a time
sharing variable \cite{Yu_06_TCOM_Dual_nonconvex_opt} in it, and
thus, $R_{2}$ is a convex $(\cap)$ function of $\rho_{0}$, and
therefore, the duality gap is zero. 

The traditional proof for the case with the constraint introduces
a time sharing variable as follows \cite{Sullivan_IT03_Information_hiding}.

\begin{eqnarray}
 &  & I(W;Y_{1}^{N},S_{2,1}^{N})\nonumber \\
 & {\le} & \sum_{n=1}^{N}I(U_{n};Y_{n},S_{2,n})-I(U_{n};S_{1,n})\label{eq:Chiang}\\
 & = & N\left(I(U;Y,S_{2}|Q)-I(U;S_{1}|Q)\right)\label{eq:add-Q}\\
 & = & N(I(U,Q;Y,S_{2})-I(Q;Y,S_{2})\nonumber \\
 &  & -I(U,Q;S_{1})+I(Q;S_{1}))\label{eq:UQ-chain}\\
 & \le & N\left(I(U,Q;Y,S_{2})-I(U,Q;S_{1})\right)\label{eq:IQS1-zero}\\
 & = & N\left(I(\bar{U};Y,S_{2})-I(\bar{U};S_{1})\right)\label{eq:newU}\\
 & \le & NR_{2}\label{eq:NR2}\end{eqnarray}
 where $W$ is the information message; $U_{n}=(W,Y_{1}^{n-1},S_{2,1}^{n-1},S_{1,n+1}^{N})$;
(\ref{eq:Chiang}) is obtained in \cite{Cover_02IT_Duality_two_sided_CSI};
(\ref{eq:add-Q}) is obtained by the definition of conditional mutual
information and by letting $Q$ be uniformly distributed over $\{1,2,...,N\}$,
$U=U_{Q}$, $S_{1}=S_{1,Q}$, $S_{2}=S_{2,Q}$, and $Y=Y_{Q}$; (\ref{eq:UQ-chain})
follows from the chain rule of the mutual information; (\ref{eq:IQS1-zero})
follows from the fact that $\{S_{1,1},...,S_{1,N}\}$ are i.i.d. and
thus, $I(Q;S_{1})=0$; (\ref{eq:newU}) follows from defining $\bar{U}=(U,Q)$;
(\ref{eq:NR2}) follows from $\frac{1}{N}\sum_{n=1}^{N}\textrm{E}\left[\alpha(X_{n})\right]=\textrm{E}\left[\alpha(X_{Q})\right]=\textrm{E}\left[\alpha(X)\right]\le\rho_{0}$
and the fact that (\ref{eq:newU}) is a convex $\cup$ function of
$p_{X|\bar{U},S_{1}}$ when $p_{\bar{U}|S_{1}}$ is fixed, which implies
that the optimal $X$ is a deterministic function of of $\bar{U}$
and $S_{1}$.

Using the Lagrangian Converse Proof, the same capacity result can
be obtained without resorting to the time sharing variable:

\begin{eqnarray}
 &  & I(W;Y_{1}^{N},S_{2,1}^{N})\nonumber \\
 & {\le} & \sum_{n=1}^{N}I(U_{n};Y_{n},S_{2,n})-I(U_{n};S_{1,n})\nonumber \\
 & \le & \sum_{n=1}^{N}\cdot\left[I(U_{n};Y_{n},S_{2,n})-I(U_{n};S_{1,n})\right.\nonumber \\
 &  & \left.-\lambda^{*}\left(\textrm{E}[\alpha(X_{n})]-\rho_{0}\right)\right],\label{eq:lag-noncausal}\\
 & \le & NC_{2},\label{eq:last-noncausal}\end{eqnarray}
 where (\ref{eq:lag-noncausal}) follows from the fact that $X_{n}$'s
satisfy the average power constraint.

So far, we have seen two examples where the duality gap is zero. One
might worry whether the proof works when the duality gap is not zero.
In the next subsection, we show that it works even when the duality
gap is not zero.

\subsection{\label{sub:feedback}Capacity of Channels with Limited Feedback and
Input Constraint}

\begin{figure}
\begin{centering}
\textsf{\includegraphics[clip,angle=270,scale=0.3]{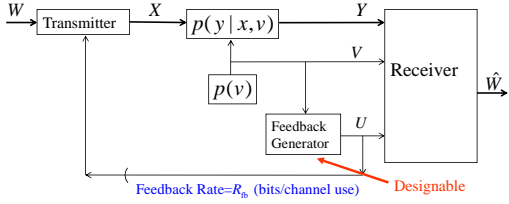}}
\par\end{centering}

\caption{\label{cap:cha-feedback} A channel with limited and designable feedback}

\end{figure}

We consider a channel with designable finite-rate/limited feedback.
As shown in Figure \ref{cap:cha-feedback}, the memoryless channel
with finite alphabets is characterized by $(\mathcal{X},\mathcal{Y},\mathcal{V},\mathcal{U},p_{V},p_{Y|X,V})$,
where $X\in{\mathcal{X}}$ is the channel input, $V\in{\mathcal{V}}$
is the channel-state with distribution $p_{V}$, $p_{Y|X,V}$ is the
channel transition probability, and $(Y\in{\mathcal{Y}},V\in{\mathcal{V}})$
is the channel output, i.e., the channel-state is know at the receiver.
For the $n^{\textrm{th}}$ channel use, the transmitter receives a
causal, but not strictly causal, finite-rate, and error free channel-state
feedback $U_{n}\in{\mathcal{U}}=\{1,...,2^{R_{\textrm{fb}}}\}$ from
the receiver. The feedback $U_{n}$ could be \emph{designed} as a
deterministic or random function of current channel-state $V_{n}$
and/or past channel-states $V_{1}^{n-1}$. Because the receiver produces
$U_{n}$, it is assumed known to the receiver. In the proof of the
converse, the inputs over $N$ channel uses satisfy the constraint,
\begin{equation}
\frac{1}{N}\sum_{n=1}^{N}\textrm{E}\left[\alpha(X_{n})\right]\le\rho_{0},\label{eq:constraint-feedback}\end{equation}
 where the expectation is over the information message and the feedback. 

The capacity \cite{Liu:IT04_SISO_feedback} of this channel without
input constraint is \begin{eqnarray}
C_{3}^{'} & = & \max_{\varphi(\cdot),p_{X|U}}I(X;Y|U=\varphi(V),V)\nonumber \\
 & = & \max_{\varphi(\cdot),p_{X|U}}\sum_{v}p(v)\nonumber \\
 &  & \cdot I\left(p_{X|U}(\cdot|\varphi(v)),\: p_{Y|X,V}(\cdot|\cdot,v)\right),\label{eq:C-Finite-noCons}\end{eqnarray}
 where the \emph{important claim} is that the feedback $U=\varphi(V)$
is a deterministic and memoryless function of the current channel-state
$V$; in (\ref{eq:C-Finite-noCons}) the mutual information is written
as a function of its input distribution and its channel transition
probability. 

Based on $C_{3}^{'}$, one might expect the capacity with input constraint
to be \begin{eqnarray}
R_{3} & = & \max_{\begin{array}{c}
\varphi(\cdot),p_{X|U}:\\
\textrm{E}[\alpha(X)]\le\rho_{0}\end{array}}I(X;Y|U=\varphi(V),V)\label{eq:primary-Finite}\end{eqnarray}
 The surprising result is that the capacity may be larger than $R_{3}$.

\begin{thm}
\label{thm:Capacity-single-finite} \cite{Liu_Asi07_Capacity_thms_feedback,Liu_IT07s_Capacity_MAC}
The capacity of the channel $(\mathcal{X},\mathcal{Y},\mathcal{V},\mathcal{U},p_{V},p_{Y|X,V})$
with designable finite-rate ($|\mathcal{U}|=2^{R_{\mbox{fb}}}$) channel-state
feedback and input constraint $\rho_{0}$ is \begin{eqnarray}
C_{3} & = & \min_{\lambda\ge0}L_{3}(\lambda,\rho_{0})\label{eq:C-Finite}\\
 & = & L_{3}(\lambda^{*},\rho_{0})\label{eq:lmd-star}\\
 & = & R_{3}+\textrm{duality gap}\nonumber \\
 & \ge & R_{3}\nonumber \end{eqnarray}
 where \begin{eqnarray}
L_{3}(\lambda,\rho_{0}) & \triangleq & \max_{\varphi(\cdot),p_{X|U}}\{I(X;Y|U=\varphi(V),V)\nonumber \\
 &  & \qquad\qquad-\lambda(\textrm{E}[\alpha(X)]-\rho_{0})\}\label{eq:dual-function}\end{eqnarray}
 is the Lagrange dual function to the primary problem (\ref{eq:primary-Finite}).
\end{thm}

\subsubsection{Without the Input Constraint}

We first review the key steps of the converse proof without the input
constraint \cite{Liu_IT07s_Capacity_MAC}. The mutual information
between the information message and the received signal is bounded
as \begin{eqnarray}
 &  & I(W;Y_{1}^{N},V_{1}^{N},U_{1}^{N})\nonumber \\
 & \le & \sum_{n=1}^{N}\sum_{u_{1}^{n-1}}p(u_{1}^{n-1})\nonumber \\
 &  & \cdot f_{3}\left(f_{1}^{(u_{1}^{n-1})}(u|v),f_{2}^{(u_{1}^{n-1})}(x|u)\right)\label{eq:nocon-f3}\\
 & \le & \sum_{n=1}^{N}\sum_{u_{1}^{n-1}}p(u_{1}^{n-1})\nonumber \\
 &  & \cdot f_{3}\left(p_{U|V}^{*}(u|v),p_{X|U}^{*}(x|u)\right)\label{eq:nocon-into-f0}\\
 & = & NC_{3}^{'},\label{eq:nocon-last}\end{eqnarray}
 where (\ref{eq:nocon-f3}) is obtained in \cite{Liu_Asi07_Capacity_thms_feedback,Liu_IT07s_Capacity_MAC}
and \begin{eqnarray*}
 &  & f_{3}\left(f_{1}(u|v),f_{2}(x|u)\right)\\
 & \triangleq & \sum_{v}p_{V}(v)\sum_{u}f_{1}(u|v)\\
 &  & \cdot I\left(f_{2}(\cdot|u),p_{Y|X,V}(\cdot|\cdot,v)\right),\end{eqnarray*}
\begin{eqnarray*}
f_{1}^{(u_{1}^{n-1})}(u|v) & = & p_{U_{n}|V_{n},U_{1}^{n-1}}(u|v,u_{1}^{n-1}),\end{eqnarray*}
\begin{eqnarray*}
f_{2}^{(u_{1}^{n-1})}(x|u) & = & p_{X_{n}|U_{n},U_{1}^{n-1}}(x|u,u_{1}^{n-1}).\end{eqnarray*}
 Let $p_{U|V}^{*}(u|v)$ and $p_{X|U}^{*}(x|u)$ be the solution to
\begin{eqnarray*}
 &  & \max_{p_{U|V}(u|v),p_{X|U}(x|u)}f_{3}\left(p_{U|V}(u|v),p_{X|U}(x|u)\right)\\
 & = & \sum_{v}p_{V}(v)\sum_{u}p_{U|V}^{*}(u|v)\\
 &  & \cdot I\left(p_{X|U}^{*}(\cdot|u),p_{Y|X,V}(\cdot|\cdot,v)\right).\end{eqnarray*}
Note that $p_{U|V}^{*}(u|v)$ and $p_{X|U}^{*}(x|u)$ are not functions
of $u_{1}^{n-1}$ because $f_{3}(\cdot,\cdot)$ is not a function
of $u_{1}^{n-1}$. Furthermore, $f_{3}\left(p_{U|V}(u|v),p_{X|U}(x|u)\right)$
is a linear function of simplex $\{p_{U|V}(u|v),u\in\mathcal{U}\}$,
and thus, the optimal $p_{U|V}^{*}(u|v)$ is obtained at the extreme
point $p_{U|V}^{*}(u|v)=\delta[u-\varphi^{*}(v)]$ for some deterministic
function $\varphi^{*}(\cdot)$, where $\delta[x]=\begin{cases}
1 & x=0\\
0 & \textrm{elsewhere}\end{cases}$. Therefore, (\ref{eq:nocon-into-f0}) and (\ref{eq:nocon-last})
are obtained.

\subsubsection{With the Input Constraint}

The traditional method reviewed in Section \ref{sub:DMC} will not
work here. One cannot produce a better feedback function and input
distribution $\left(p_{U|V}(u|v),p_{X|U}(x|u)\right)$ by averaging
$\left(f_{1}^{(u_{1}^{n-1})}(u|v),f_{2}^{(u_{1}^{n-1})}(x|u)\right)$
because $f_{3}\left(p_{U|V}(u|v),p_{X|U}(x|u)\right)$ is not a convex
function of $\left(p_{U|V}(u|v),p_{X|U}(x|u)\right)$. However, one
could introduce a time sharing variable, as shown in Section \ref{sub:non-causal},
but the time sharing variable cannot be absorbed into an existing
auxiliary variable of the capacity formula as in (\ref{eq:newU}). 

Therefore, we resort to the Lagrangian Converse Proof \cite{Liu_IT07s_Capacity_MAC}.
The key steps are 

\begin{eqnarray}
 &  & I(W;Y_{1}^{N},V_{1}^{N},U_{1}^{N})\nonumber \\
 & \le & \sum_{n=1}^{N}\sum_{u_{1}^{n-1}}p(u_{1}^{n-1})\nonumber \\
 &  & \cdot f_{3}\left(f_{1}^{(u_{1}^{n-1})}(u|v),f_{2}^{(u_{1}^{n-1})}(x|u)\right)\nonumber \\
 &  & -\lambda^{*}\sum_{n=1}^{N}\left(\textrm{E}\left[\alpha(X_{n})\right]-\rho_{0}\right)\label{eq:feedback-lambda}\\
 & \le & \sum_{n=1}^{N}\sum_{u_{1}^{n-1}}p(u_{1}^{n-1})\nonumber \\
 &  & \cdot f_{4}\left(p_{U|V}^{*}(u|v),p_{X|U}^{*}(x|u),\lambda^{*}\right)\label{eq:feedback-f4}\\
 & = & NC_{3},\label{eq:feedback-last}\end{eqnarray}
 where $\lambda^{*}$ is the solution to (\ref{eq:lmd-star}); (\ref{eq:feedback-lambda})
follows from the fact that the constraint is satisfied and thus $-\lambda^{*}\sum_{n=1}^{N}\left(\textrm{E}\left[\alpha(X_{n})\right]-\rho_{0}\right)\ge0$;
and 

\begin{eqnarray*}
 &  & f_{4}\left(f_{1}(u|v),f_{2}(x|u),\lambda\right)\\
 & \triangleq & \sum_{v}p_{V}(v)\sum_{u}f_{1}(u|v)\\
 &  & \cdot\left[I\left(f_{2}(\cdot|u),p_{Y|X,V}(\cdot|\cdot,v)\right)\right.\\
 &  & \left.-\lambda\left(\sum_{x}f_{2}(x|u)\alpha(x)-\rho_{0}\right)\right].\end{eqnarray*}
 Let $p_{U|V}^{*}(u|v)$ and $p_{X|U}^{*}(x|u)$ be the solution to
\begin{eqnarray*}
 &  & \max_{p_{U|V}(u|v),p_{X|U}(x|u)}f_{4}\left(p_{U|V}(u|v),p_{X|U}(x|u),\lambda^{*}\right).\end{eqnarray*}
 Again, because $f_{4}(\cdot,\cdot,\cdot)$ is not a function of $u_{1}^{n-1}$
and $f_{4}\left(p_{U|V}(u|v),p_{X|U}(x|u),\lambda^{*}\right)$ is
a linear function of the simplex $\{p_{U|V}(u|v),u\in\mathcal{U}\}$,
one obtains that $p_{U|V}^{*}(u|v)=\delta[u-\varphi^{*}(v)]$ and
$p_{X|U}^{*}(x|u)$ are not functions of $u_{1}^{n-1}$. Therefore,
(\ref{eq:feedback-f4}) and (\ref{eq:feedback-last}) are obtained.

\subsubsection{Relation of the Lagrange Dual Function to the Time Sharing and the
Capacity Region}

In the following, we illustrates the central role of the Lagrange
dual function $L_{3}$ from two aspects.

\subsubsection*{Time Sharing}

We first discuss a time sharing expression $C_{3}^{\textrm{TS}}$
of the capacity and then show that $C_{3}^{\textrm{TS}}=C_{3}$ using
the Lagrange dual function $L_{3}$. The alternative converse proof
using time sharing is as follows. Define the random variable $Q_{1}$
to be uniformly distributed over $\{1,...,N\}$ and another one to
be $Q_{2}=U_{1}^{Q_{1}-1}$. Then define the time sharing random variable
$Q\triangleq(Q_{1},Q_{2})\in\mathcal{Q}$. We obtain \begin{eqnarray}
 &  & I(W;Y_{1}^{N},V_{1}^{N},U_{1}^{N})\nonumber \\
 & \le & \sum_{n=1}^{N}\sum_{u_{1}^{n-1}}p(u_{1}^{n-1})\nonumber \\
 &  & \cdot f_{3}\left(f_{1}^{(u_{1}^{n-1})}(u|v),f_{2}^{(u_{1}^{n-1})}(x|u)\right)\nonumber \\
 & = & N\sum_{q}p_{Q}(q)\sum_{v}p(v)\sum_{u}p_{U|V,Q}(u|v,q)\nonumber \\
 &  & \cdot I\left(p_{X|U,Q}(\cdot|u,q),p_{Y|X,V}(\cdot|\cdot,v)\right)\nonumber \\
 & = & NI(X;Y|U,V,Q)\label{eq:NIUVQ}\\
 & \le & NC_{3}^{\textrm{TS}},\label{eq:time-last}\end{eqnarray}
 where \begin{eqnarray}
C_{3}^{\textrm{TS}} & = & \max_{\mathcal{Q},p_{Q},\varphi_{Q}(\cdot),p_{X|U,Q}:\textrm{E}[\alpha(X)]\le\rho_{0}}\nonumber \\
 &  & I(X;Y|U=\varphi_{Q}(V),V,Q);\label{eq:cap-time-share}\end{eqnarray}
 and (\ref{eq:time-last}) follows from the fact that (\ref{eq:NIUVQ})
is a linear function of the simplex $\{p_{U|V,Q}(u|v,q),u\in\mathcal{U}\}$
and thus the deterministic feedback $U=\varphi_{Q}(V)$ does not lose
the optimality. 

It turns out that the Lagrange dual function $L_{3}$ in (\ref{eq:dual-function})
is not only the dual to the primary problem $R_{3}$ in (\ref{eq:primary-Finite}),
but also the dual to the optimization of $C_{3}^{\mbox{TS}}$ in (\ref{eq:cap-time-share}):
\begin{eqnarray}
L_{3}^{\textrm{TS}}(\lambda,\rho_{0}) & \triangleq & \max_{\mathcal{Q},p_{Q},\varphi_{Q}(\cdot),p_{X|U,Q}}\sum_{q\in\mathcal{Q}}p_{Q}(q)\cdot\nonumber \\
 &  & \{I(X;Y|U=\varphi(V),V,Q=q)\nonumber \\
 &  & -\lambda(\textrm{E}[\alpha(X)|Q=q]-\rho_{0})\}\label{eq:dual-function-time}\\
 & = & L_{3},\label{eq:get-to-L3}\end{eqnarray}
 where (\ref{eq:get-to-L3}) follows the fact that the function to
be optimized in (\ref{eq:dual-function-time}) is a linear function
of the simplex $\{p_{Q}(q),q\in\mathcal{Q}\}$ and thus, the optimal
solution is obtained at certain $q^{*}$ for which $p_{Q}(q^{*})=1$.
Therefore, the one dual function for two primary problems shows that
$C_{3}^{\textrm{TS}}=C_{3}$.

\subsubsection*{Capacity Regions}

We show that the Lagrange dual function $L_{3}$ characterizes the
boundary points of the two expressions, $\mathcal{C}_{3}^{\textrm{TS}}$
and $\mathcal{C}_{3}$, of the single user capacity region. Equation
(\ref{eq:NIUVQ}) shows that any achievable rate $r$ under constraint
$\rho$ must belong to the following capacity region: \begin{eqnarray}
\mathcal{C}_{3}^{\textrm{TS}} & = & \mbox{closure}\bigcup_{\mathcal{Q},p_{Q},p_{U|V,Q},p_{X|U,Q}}\nonumber \\
 &  & \mathcal{C}_{3,\textrm{Fixed}}^{\textrm{TS}}\left(\mathcal{Q},p_{Q},p_{U|V,Q},p_{X|U,Q}\right),\label{eq:c-region-TS}\end{eqnarray}
 where\begin{eqnarray}
 &  & \mathcal{C}_{3,\textrm{Fixed}}^{\textrm{TS}}\left(\mathcal{Q},p_{Q},p_{U|V,Q},p_{X|U,Q}\right)\nonumber \\
 & = & \left\{ (r,\rho):0\le r\le I(X;Y|U,V,Q),\textrm{E}[\alpha(X)]\le\rho\right\} .\label{eq:c-region-fixed-TS}\end{eqnarray}
 Note that following the leads by Gallager in the study of non-convex
multiple access capacity region \cite{Gallager_Rpt87_Energy_limited_channels},
we have included $\rho$ to make the capacity region a two dimensional
set. Since a convex hull performs the time sharing for you, an equivalent
capacity region is \begin{eqnarray}
\mathcal{C}_{3} & = & \mbox{closure convex}\bigcup_{p_{U|V},p_{X|U}}\nonumber \\
 &  & \mathcal{C}_{3,\textrm{Fixed}}\left(p_{U|V},p_{X|U}\right)\label{eq:c-region}\\
 & = & \mathcal{C}_{3}^{\textrm{TS}},\nonumber \end{eqnarray}
 where\begin{eqnarray}
 &  & \mathcal{C}_{3,\textrm{Fixed}}\left(p_{U|V},p_{X|U}\right)\nonumber \\
 & = & \left\{ (r,\rho):0\le r\le I(X;Y|U,V),\textrm{E}[\alpha(X)]\le\rho\right\} .\label{eq:c-region-fixed}\end{eqnarray}

Characterizing the boundary of $\mathcal{C}_{3}^{\textrm{TS}}$ and
$\mathcal{C}_{3}$ can be reduced to solving the Lagrange dual function
$L_{3}$. Let $(1,-\lambda)$ be the normal vector of a hyperplane.
Finding the points of $\mathcal{C}_{3}^{\textrm{TS}}$ that touch
the hyperplane needs to solve \begin{eqnarray*}
B_{3}^{\textrm{TS}}(\lambda) & \triangleq & \max_{(r,\rho)\in\mathcal{C}_{3}^{\textrm{TS}}}(1,-\lambda)\cdot(r,\rho)\\
 & = & \max_{(r,\rho_{})\in\mathcal{C}_{3}^{\textrm{TS}}}r-\lambda\rho_{},\end{eqnarray*}
 which can be reduced to \begin{eqnarray*}
r & = & I(X;Y|U,V,Q)\\
\rho_{} & = & \textrm{E}[\alpha(X)]\\
B_{3}^{\textrm{TS}}(\lambda) & = & L_{3}^{\textrm{TS}}(\lambda,\rho_{0})+\lambda\rho_{0}\\
 & = & L_{3}(\lambda,\rho_{0})+\lambda\rho_{0}.\end{eqnarray*}
 The same is true for $\mathcal{C}_{3}$: \begin{eqnarray*}
r & = & I(X;Y|U,V)\\
\rho_{} & = & \textrm{E}[\alpha(X)]\\
B_{3}(\lambda) & = & L_{3}(\lambda,\rho_{0})+\lambda\rho_{0}.\end{eqnarray*}

Therefore, we have seen that the Lagrange dual function plays the
central role to connect the boundary points of the capacity region
and the capacity expressions: \begin{eqnarray*}
B_{3}^{\textrm{TS}}(\lambda)-\lambda\rho_{0} & = & L_{3}^{\textrm{TS}}(\lambda,\rho_{0})\\
=B_{3}(\lambda)-\lambda\rho_{0} & = & L_{3}(\lambda,\rho_{0})\\
 & \ge & \min_{\lambda\ge0}L_{3}(\lambda,\rho_{0})\\
 & = & C_{3}(\rho_{0})=C_{3}^{\textrm{TS}}(\rho_{0})\\
 & \ge & R_{3}(\rho_{0}).\end{eqnarray*}

\begin{remrk}
Expressing the capacity as the minimum of the Lagrange dual function
also helps to calculate the capacity because one does not need to
worry about the time sharing while performing the optimization. If
multiple solutions, i.e., input distributions etc., achieve the same
value of the Lagrange dual function, then the capacity achieving strategy
is a time sharing of these solutions and the time sharing coefficients
are chosen to satisfy the constraint. See \cite{Liu_Asi07_Capacity_thms_feedback,Liu_IT07s_Capacity_MAC}
for details.
\end{remrk}
\begin{figure}
\begin{centering}
\textsf{\includegraphics[clip,scale=0.65]{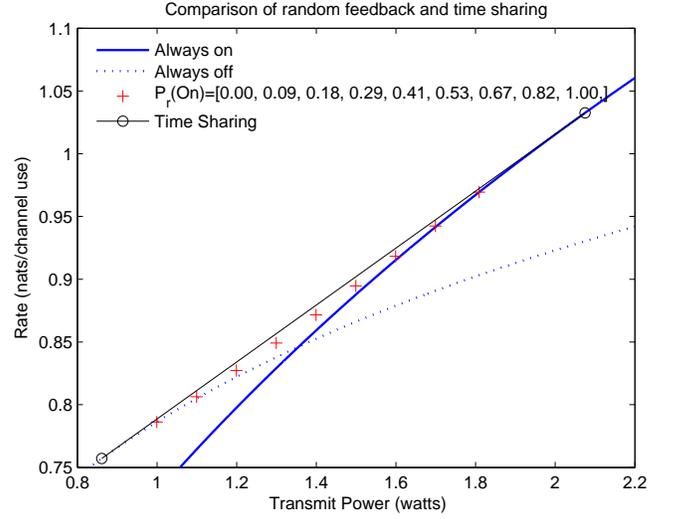}}
\par\end{centering}

\caption{\label{cap:rand-feedback} An example of nonzero duality gap.}

\end{figure}

\begin{example}
To illustrate the capacity with nonzero duality gap, we produced an
example, whose detailed derivation is given in \cite{Liu_Asi07_Capacity_thms_feedback,Liu_IT07s_Capacity_MAC}.
The channel is an additive Gaussian noise channel with three states,
good, moderate, and bad states, corresponding to small, moderate,
and large noise variances. The feedback is limited to 1 bit/channel
use. For small long term average power constraint, the optimal strategy
is to turn on the transmitter with a fixed power only when the channel
is in the good state, as shown by the dotted curve in Figure \ref{cap:rand-feedback}.
For large power constraint, the optimal strategy is to turn on the
transmitter when the channel is in good or moderate state with another
fixed power, as shown by the solid curve in Figure \ref{cap:rand-feedback}.
For the power constraint in between, the optimal strategy is a time
sharing of the above two strategies, as shown by the line segment
terminated by the ``o''s. The gap between the line segment and the
maximum of the dotted and the solid curves is exactly the nonzero
duality gap between $C_{3}$ and $R_{3}$. The slope of the line segment
is $\lambda^{*}$. The ``+'' markers are for random feedback discussed
in \cite{Liu_Asi07_Capacity_thms_feedback,Liu_IT07s_Capacity_MAC}.
\end{example}

\section{The Extension to the Rate Distortion Theory\label{sec:rate-distortion}}

\subsection{The Converse Proof}

It is straight forward to extend the Lagrangian Converse Proof to
the rate distortion theory. We illustrate it using the classic i.i.d.
source as an example. The rate distortion function of quantizing i.i.d.
source $X$ to $\hat{X}$ in a vector manner is \begin{eqnarray*}
R'_{1}(D) & = & \min_{p_{\hat{X}|X}:\mbox{E}[d(X,\hat{X})]\le D}I(X;\hat{X}),\end{eqnarray*}
 where $d(\cdot,\cdot)$ measures the distortion. Use the Lagrange
dual function, we have another expression \begin{eqnarray}
R_{1}(D) & = & \max_{\lambda\ge0}L_{1}(\lambda,D),\label{eq:lag-expression}\end{eqnarray}
 where

\begin{eqnarray*}
L_{1}(\lambda,D) & \triangleq & \min_{p_{\hat{X}|X}}I(X;\hat{X})+\lambda\left(\mbox{E}[d(X,\hat{X})]-D\right).\end{eqnarray*}
 In general, the Lagrange dual function is a lower bound and we have
$R_{1}(D)\le R'_{1}(D)$. Due to the convexity of the mutual information,
we have $R_{1}(D)=R'_{1}(D)$.

The last few steps of the conventional converse proof is \cite{Cover:91}
\begin{eqnarray}
 &  & \sum_{n=1}^{N}I(X_{n};\hat{X}_{n})\nonumber \\
 & \ge & \sum_{n=1}^{N}R'_{1}(\mbox{E}[d(X_{n},\hat{X}_{n})])\nonumber \\
 & \ge & nR'_{1}\left(\frac{1}{n}\sum_{n=1}^{N}\mbox{E}[d(X_{n},\hat{X}_{n})]\right)\label{eq:R-convex}\\
 & = & nR'_{1}(D),\nonumber \end{eqnarray}
 where (\ref{eq:R-convex}) used the property that $R'_{1}(D)$ is
a convex $\cup$ function of $D$. 

The Lagrangian Converse Proof does not need to prove the the convexity
property of $R'_{1}(D)$ before performing the converse proof: \begin{eqnarray}
 &  & \sum_{n=1}^{N}I(X_{n};\hat{X}_{n})\nonumber \\
 & \ge & \sum_{n=1}^{N}\left(I(X_{n};\hat{X}_{n})+\lambda^{*}\left(\mbox{E}[d(X_{n},\hat{X}_{n})]-D\right)\right)\label{eq:lag-rate-dist}\\
 & \ge & nR_{1}(D),\label{eq:last-rate-dist}\end{eqnarray}
 where $\lambda^{*}\ge0$ is the solution to (\ref{eq:lag-expression});
(\ref{eq:lag-rate-dist}) follows from the fact that the distortion
requirement is satisfied by $\hat{X}_{n}$'s and thus $\lambda^{*}\left(\left(\sum_{n=1}^{N}\mbox{E}[d(X_{n},\hat{X}_{n})]\right)-ND\right)\le0$;
(\ref{eq:last-rate-dist}) follows from the fact that $R_{1}(D)$
lower bound the summand in (\ref{eq:lag-rate-dist}) for \emph{every}
$n$.

The benefit of the Lagrangian Converse Proof may not appear to be
significant in this simple example. But it can be easily applied to
more complex cases when the time sharing has to be used in $R'_{1}(D)$.
Another example is when there are other constraints in addition to
the distortion, in which case, simply introducing more Lagrange multipliers
solves the problem.

\subsection{Dual Relation between Channel Capacity and Rate Distortion}

\begin{figure*}[!t]
\begin{eqnarray}
C_{2}(\rho_{0}) & = & \min_{\lambda\ge0}\max_{\mathcal{U},X=\varphi(U,S_{1}),p_{U|S_{1}}}I(U;S_{2},Y)-I(U;S_{1})-\lambda(\textrm{E}[\alpha(X)]-\rho_{0})\label{eq:c2-all}\\
R_{2}(D) & = & \max_{\lambda\ge0}\min_{\mathcal{U},\hat{X}=f(U,S_{2}),p_{U|X,S_{1}}}I(U;S_{1},X)-I(U;S_{2})+\lambda(\textrm{E}[d(X,\hat{X})]-D),\label{eq:r2-all}\end{eqnarray}
\lyxline{\normalsize}
\end{figure*}

We note that using expressions involving Lagrange dual functions,
the channel capacity and the rate distortion function has a pleasant
symmetric form, as evident in $C_{1}(\rho_{0})$ (\ref{eq:C-Finite-no-feedback})
and $R_{1}(D)$ (\ref{eq:lag-expression}) for channels without side
information. The symmetric form shows a dual relation in the sense
of \cite{Cover_02IT_Duality_two_sided_CSI}. 

It can be easily extended to the case of non-causal side information
considered in \cite{Cover_02IT_Duality_two_sided_CSI}, where the
constraints of the capacity is not considered. With the constraint,
the capacity (\ref{eq:c2-all}) and the rate distortion (\ref{eq:r2-all})
are shown at the top of the next page. The dual relation defined in
\cite{Cover_02IT_Duality_two_sided_CSI} is the following isomorphism.
\begin{eqnarray*}
\textrm{Channel Capacity} &  & \textrm{Rate Distortion}\\
\min & \longleftrightarrow & \max\\
\max & \longleftrightarrow & \min\\
-\lambda & \longleftrightarrow & +\lambda\\
\textrm{Transmitted Symbol }X & \longleftrightarrow & \hat{X}\textrm{ Estimation}\\
\textrm{Received Symbol }Y & \longleftrightarrow & X\textrm{ Source}\\
\textrm{State to Encoder }S_{1} & \longleftrightarrow & S_{2}\textrm{ State to Decoder}\\
\textrm{State to Decoder }S_{2} & \longleftrightarrow & S_{1}\textrm{ State to Encoder}\\
\textrm{Auxiliary }U & \longleftrightarrow & U\textrm{ Auxiliary}\\
\textrm{Input Cost }\alpha(\cdot) & \longleftrightarrow & d(\cdot,\cdot)\textrm{ Distortion Measure}\\
\textrm{Input Constraint }\rho_{0} & \longleftrightarrow & D\textrm{ Distortion}.\end{eqnarray*}

A stronger dual relation is defined in \cite{Pradhan_IT03_Duality_source_channel_side},
where the capacity and the rate distortion can be made equal by selecting
proper constraints. But it does not work when the optimal solutions
need time sharing. Since (\ref{eq:c2-all}) and (\ref{eq:r2-all})
do not include the time sharing variables, it is a future research
to see whether the stronger dual relation can be established with
some modification.

The dual relation for the limited feedback case is not discussed here.
The reason is that the not-strictly-causal feedback to the transmitter
in channel capacity corresponds to finite rate state information to
the decoder in rate distortion. While the encoder in channel capacity
cannot use future feedback, the decoder in rate distortion can wait
to use both past and future finite rate state information.

\section{Conclusions\label{sec:Conclusions}}

We have introduced a simple converse proof that uses the Lagrange
dual function to upper bound the information rate. It provides the
following approach to deal with constraints: 1) Based on the capacity
of the channel without constraints, express the capacity for the case
with the constraints as the minimum of the Lagrange dual function;
2) Simply modify the converse proof for the case without the constraints
by adding to the second to the last expression a term involving the
Lagrange multiplier and the constraints, to produce the converse proof
for the case with the constraints; 3) For the achievability, study
the duality gap to determine whether the time sharing is needed. 

We show that the unified capacity expression, \begin{eqnarray*}
C & = & \min_{\lambda\ge0}\textrm{ Lagrange Dual Function}(\lambda),\end{eqnarray*}
 plays a central role to connect the characterization of the single
user capacity region, the time sharing capacity formula, and the formula
resulted by imposing the constraint to the maximization in the capacity
formula of the case without constraints. The Lagrangian capacity formula
works regardless whether the problem is convex or not. This formula
also simplifies the evaluation of the capacity, by deferring the consideration
of the time sharing.

The above is extended to the rate distortion theory. A symmetric form
of capacity and rate distortion function is shown to demonstrate the
dual relation between them. Further extension to the case of multiple
constraints is straight forward. We have discussed the single letter
capacity formula in this paper. The extension of the Lagrangian Converse
Proof to multi-letter capacity formula, multiaccess channels, and
broadcast channels is deferred to future research.

\bibliographystyle{IEEEtran}
\bibliography{liu_all}

\end{document}